\documentclass[twocolumn]{aastex63}
\usepackage{graphicx}
\usepackage{float}

\submitjournal{ApJ}
\shorttitle{HST Proper Motion of Andromeda~III}
\shortauthors{Casetti-Dinescu et al.}

\begin{document}

\title{HST Proper Motion of Andromeda~III: Another Satellite
Co-orbiting The M31 Satellite Plane}

\correspondingauthor{Dana I. Casetti-Dinescu}
\email{casettid1@southernct.edu,dana.casetti@gmail.com}

\author[0000-0001-9737-4954]{Dana I. Casetti-Dinescu}
\affiliation{Department of Physics, Southern Connecticut
  State University, 501 Crescent Street, 
New Haven, CT 06515, USA}
\affiliation{Astronomical Institute of the
  Romanian Academy, Cutitul de Argint 5, Sector 4, 
Bucharest, Romania}
\author[0000-0002-9197-9300]{Marcel S. Pawlowski}
\affiliation{Leibnitz-Institut f\"{u}r Astrophysik (AIP),
  And der Sternwarte 16, D-14482 Potsdam, Germany}
\author[0009-0001-3739-7051]{Terrence M. Girard}
\affiliation{Department of Physics, Southern Connecticut
  State University, 501 Crescent Street, New Haven, CT 06515, USA}
\author[0000-0002-2483-2595]{Kosuke Jamie Kanehisa}
\affiliation{Leibnitz-Institut f\"{u}r Astrophysik (AIP),
  And der Sternwarte 16, D-14482 Potsdam, Germany}
\author{Alexander Petroski}
\affiliation{Department of Physics, Southern Connecticut
  State University, 501 Crescent Street, 
  New Haven, CT 06515, USA}
\author{Max Martone}
\affiliation{Department of Physics, Southern Connecticut
  State University, 501 Crescent Street, 
  New Haven, CT 06515, USA}
\author[0000-0003-0218-386X]{Vera Kozhurina-Platais}
\affiliation{Space Telescope Science Institute, 3700 San Martin Drive, Baltimore, MD 21218, USA}
\affiliation{Eureka Scientific, Inc., 2452 Delmer Street, Suite 100, Oakland, CA 94602-3017, USA}
\author[0000-0003-2599-2459]{Imants Platais}
\affiliation{Department of Physics and Astronomy, Johns Hopkins
  University, 3400 North Charles Street, 
  Baltimore, MD 21218, USA}

\begin{abstract}

  We measure the absolute proper motion of Andromeda~III (And\,III)
  using ACS/WFC and WFPC2 exposures spanning an
  unprecedented $22$-year time baseline.
  The WFPC2 exposures have been processed using a deep-learning
  centering procedure recently developed as well as an
  improved astrometric calibration of the camera.
  The absolute proper motion zero point is given by
  98 galaxies and 16 {\it Gaia} EDR3 stars. 
  The resulting proper motion is
  $(\mu_{\alpha} , \mu_{\delta}) = (-10.5\pm12.5, 47.5\pm12.5)~\mu$as yr$^{-1}$.
  We perform an orbit analysis of And\,III using
  two estimates of M31's mass and proper motion. We find that 
  And\,III's orbit is consistent with
  dynamical membership to the Great Plane of Andromeda
  system of satellites although with some looser alignment
  compared to the previous two satellites NGC\,147 and NGC\,185.
  And\,III is bound to M31 if M31's mass is
  $M_{\mathrm{vir}}\geq 1.5\times10^{12}\,M_{\odot}$.
  
\end{abstract}

\keywords{Astrometry: Space astrometry --- Proper motions: --- Andromeda Galaxy:
--- Dwarf elliptical galaxies: --- Local Group:}

\section{Introduction \label{sec:intro}}

Precision astrometry is revolutionizing our
understanding of the local universe, all
the way from exoplanets to
the Milky Way and the Andromeda systems.
Proper motions are the crucial measurements
in deriving 3D velocities,
which in turn open the access
to a full dynamical analysis of a given system.
Currently, this is only possible 
for the Milky Way galaxy  and its satellite system,
and for Andromeda (M31) and some of its satellites.
Large distances to these systems imply very
small tangential (angular) velocities which
in turn require proper-motion uncertainties
at the level of tens to a few of $\mu$as~yr$^{-1}$.

Active effort in this direction requires
primarily space-based observatories
such as {\it HST, JWST} and {\it Gaia}.
Unfortunately, {\it Gaia} cannot reach to
the distances needed for directly measuring
faint systems at $\sim 100$ kpc or more, at least not
at the required precision for dynamical studies.
On the other hand, {\it HST} and {\it JWST} are
adequate platforms, provided time baselines
in excess of 10 years are available.
In which case {\it Gaia} can help to calibrate 
such {\it HST}-based
studies, as we demonstrate in this study
\citep[see also][]{casetti2022,bennet2023,warfield2023}.

Much like our own Milky Way system, the Andromeda
system of satellites has about half
of its satellites in a narrow plane 
referred to as the Great Plane of Andromeda (GPoA)
\citep{conn2013,savino2022}.
From line-of-sight velocities and the geometry of the
plane, it is known that these satellites have a
coherent motion suggesting a rotation in this plane
\citep{ibata2013}.

The origin and evolution of such thin, 
rotationally supported satellite configurations
is currently unsettled; however, it is clear
that these structures are not easily formed in
state-of-the-art cosmological simulations, if at all.
\citet{taibi2024}, for instance, discuss in detail
the Milky-Way system and highlight the astonishing
degree of phase-space correlations for such in-plane
satellites.

Regarding the M31 system, only two
in-plane satellites have recently been measured
by \citet{sohn2020}. Specifically, NGC\,147 and
NGC\,185 were found to have proper motions
consistent with co-rotation in this plane
\citep{sohn2020,ps2021}.

In this paper we present the measurement of
And\,III, a satellite situated on the southern side of
M31, opposite NGC\,147 and NGC\,185, and only a little bit
off the GPoA. Its line-of-sight velocity relative to
M31 indicates And\,III is approaching us, while
NGC\,147 and NGC\,185 are receding. We also present an
orbit analysis of And\,III and argue in favor of its
membership to the GPoA. 

\section{Data Sets \label{sec:set}}
Our proper-motion study is based on HST observations from three epochs,
namely 1999, 2014 and 2021. The earliest
epoch consists of WFPC2 exposures while the later ones are
ACS/WFC exposures.
All image data were downloaded from the Mikulski Archive for
Space Telescopes (MAST).
Details of these exposures
are given in Table \ref{tab:prop-data}, including the original observing
proposal id number (PID) for reference. All exposures are
well-centered on And III, with relatively
small offsets of between a few tenths of an arcsec to
at most 3 arcsec.
Astrometrically, the whole of the data set is of high value for two reasons:
1) the long time baseline of over 22 years lessens the impact of any residual,
uncorrected positional errors (both systematic and random) within each detector, 
enhancing the true, proper-motion signal of our target, and
2) the existence of mid-epoch data serves as a useful tool for detecting any 
remaining systematic errors between the extreme-epoch data, as well as providing 
a check on spurious, outlier proper-motion values for individual stars.

\begin{figure}
    \centering
    \includegraphics[scale=0.45,angle=-90]{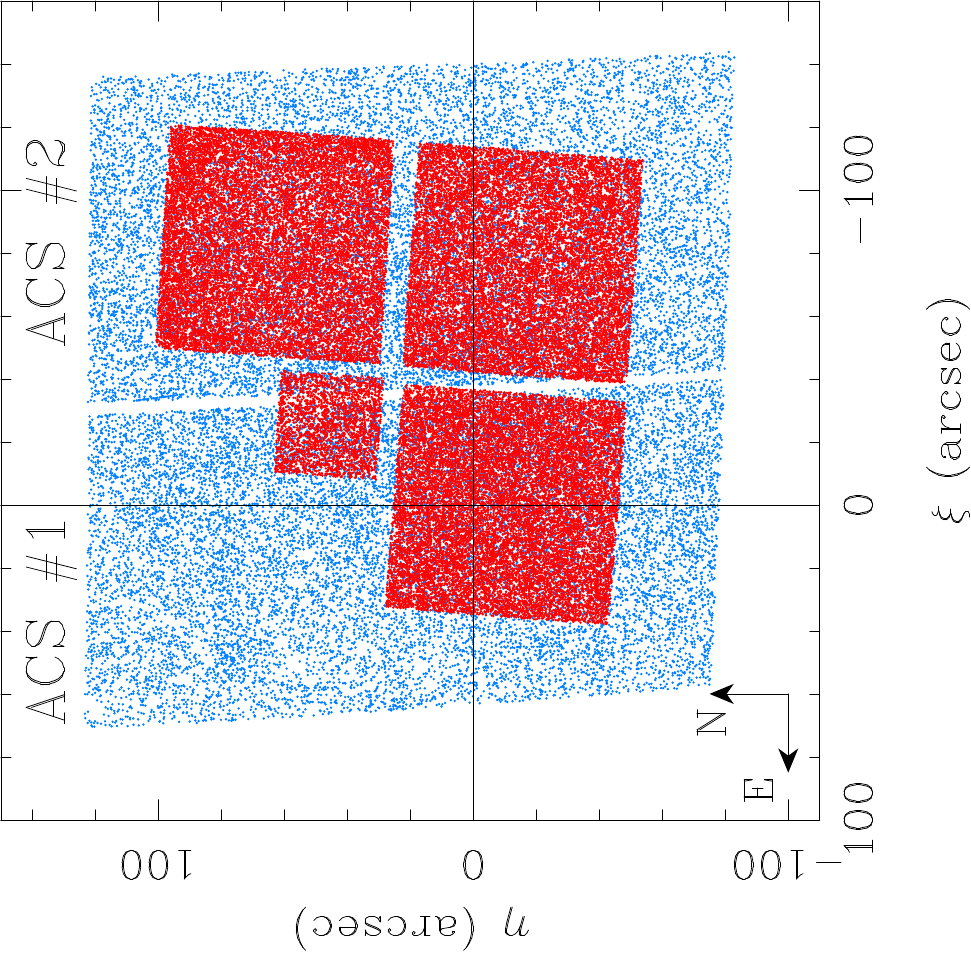}
    \caption{Overlap between the 1999 WFPC2 exposures (red symbols) and
      the modern 2014/2021 ACS/WFC exposures (blue symbols).
      Coordinates are a gnomonic projection of RA and DEC
      with the tangent point $(\xi, \eta) = (0,0)$ at
      And\,III's center
      \citep{mcconn2012}.}
    \label{fig:overlap}
\end{figure}

In Figure \ref{fig:overlap} we illustrate the
typical overlap between the
earliest 1999 WFPC2 exposures and the later ACS/WFC exposures.
The figure actually shows a 2014 ACS exposure, however, the
2021 ACS exposures were designed to align with the 2014 ones.

The center of And III, in equatorial coordinates, is adopted from
\citet{mcconn2012}. 
In Fig. \ref{fig:overlap}
$(\xi, \eta)$ are the
gnomonic projection of the equatorial coordinate system
about this position.
That is, the nominal center of And\,III is at $(\xi, \eta) = (0, 0)$
in the figure.

\begin{deluxetable}{cccr}
  \tablecaption{Properties of the Data Sets
    \label{tab:prop-data}}
\tablewidth{0pt}
\tablehead{
    \colhead{Camera} &
    \colhead{Epoch} &
    \colhead{Exposures} &
    \colhead{PID} 
}
\startdata
WFPC2    & 1999.15 & $16\times1300~$sec - F450W & 7500 \\
WFPC2    & 1999.15 & $8\times1200~$sec - F555W &  7500 \\ \\
ACS/WFC  & 2014.90 & $11\times1264~$sec; $11\times1372$~sec - F475W & 13739 \\ 
ACS/WFC  & 2014.90 & $11\times1086~$sec; $11\times1002$~sec - F814W & 13739 \\ \\
ACS/WFC  & 2021.93 & $7\times60~$sec; $14\times1118$~sec - F814W & 16273 \\
\enddata
\end{deluxetable}

\section{Data Processing \label{sec:proc}}
\subsection{WFPC2 \label{subsec:wfpc2}}
The WFPC2 standard-calibrated $_{-}$c0m.fits images
from MAST are first
corrected for cosmic-ray (CR) contamination. 
The purpose of the CR-cleaning is to rectify the sky background statistics
upon which the source-detection threshold depends, as well as to prevent the
CRs themselves from being mistakenly detected as sources.
Pairs of exposures with offsets less than 1 pixel are used to replace pixel
values in the target image with those from the comparison image,
when the difference in pixel values exceeds a given value-dependent level.

The CR-cleaned images are then processed for source detection
and first-estimate centering
using the classic {\it effective} PSF (ePSF) algorithm, i.e. the {\it hst1pass} code, 
2023 version \citep{and2000,and2022}. We note that WFPC2 ePSF libraries for
filter F450W do not exist, thus we have used the F555W libraries for such images.

The next crucial step is refinement of the source $(x,y)$ centers by
application of the
deep-learning (DL) model recently developed by 
\citet{casetti2024}. This DL model was trained on a large set
of WFPC2 exposures, also taken in 1999, at the core of
globular cluster 47 Tuc. This latest DL model is an improvement over
an earlier version \citep{casetti2023} in that it
takes into account the PSF variation across the chip and as a function of
magnitude. It has been developed for two filters, namely F555W and
F814W. Here, we use the F555W model for both the
F450W and F555W And\,III exposures.

It is worth emphasizing that
WFPC2 images are severely undersampled and thus
suffer from a fractional-pixel bias
in the stars' calculated centers, the result of any mismatch between the actual PSF 
and that assumed by the centering algorithm.
In our experience, the standard hst1pass ePSF library for WFPC2
does not adequately remove this pixel-phase bias in many cases.
The amplitude of the residual bias can
as large as 40 to 50 mpix \citep{casetti2024}.
Unfortunately, once other astrometric corrections are applied,
this bias is hard to detect as such, and essentially manifests itself as
additional random noise in the stars' centers.
Our DL technique has proved to be successful on a number of other
cluster targets \citep{casetti2024}, therefore
we choose to utilize it here.
Here, we evaluated the impact of using DL centers versus {\it hst1pass} ones,
utilizing the ``bias curve'' tool 
developed in \citet{casetti2021}, with which the distribution of
the fractional pixel for all centers is compared to that
of a uniform distribution (see e.g., Fig. 3 in
\citet{casetti2021}). 
The result, for these And\,III WFPC2 exposures, is that DL centering
improves the residual bias, compared to {\it hst1pass} centering,
by a factor of two, decreasing it roughly to the random noise level.

Once raw pixel positions are in hand, the following corrections are made:
the well-known 34th-row correction \citep{and1999},
classic 3rd-order distortion correction \citep{and2003}, followed by the
more recent higher-order distortion corrections mapped by \citet{casetti2021}. 
These distortion corrections were developed for filters other
than F450W, thus, for the images in this filter, we have used the corrections for
the closest-wavelength filter, namely F555W.


To assess the precision of the positions thus obtained,
we perform a ``plate'' transformation between a
reference exposure and a target exposure.
The transformation is a classic polynomial one that includes up to
3rd-order terms \citep[see e.g.,][]{casetti2021}, and the
scatter of the residuals represents the errors in position
of both the reference and the target image.
Limiting the analysis to well-measured stars, we obtain single-measurement standard
errors of 44 mpix for the PC and 36 mpix for the WF chips, in filter F555W.
This corresponds to 2 mas for the
PC and 3.6 mas for the WF. These positional uncertainties
are much larger than what we typically obtain for the WFPC2 camera
in cluster fields, where we achieve errors of the order of
10 mpix \citep{casetti2024}. The discrepancy is primarily attributable to
an issue of signal-to-noise: the And\,III images we are measuring are
predominantly faint stars that --- in spite of the long exposure time ---
are in a lower signal-to-noise regime than the stars in more nearby
globular clusters. In addition, CRs, although corrected for, may
still affect the centering for a fraction of the stars.
For filter F450W,
positional errors are only a few percent larger than those 
for F555W, which is notable, considering the astrometric
corrections we applied were developed for F555W.

\subsection{ACS/WFC \label{subsec:acs}}

For the ACS/WFC data, we work with $_{-}$flt.fits files from MAST that have undergone
the standard HST-pipeline calibration for bias, dark, and flat-field corrections.
With these files, no charge transfer efficiency (CTE)
correction has been applied to the image data. Rather, the CTE correction will be
that implemented internally by {\it hst1pass}.
CTE-corrected image data are available from MAST, i.e., $_{-}$flc-type fits files,
and we have performed centering tests using these files as well.
For the short ACS exposures, (see Tab. \ref{tab:prop-data}), 
we find the $_{-}$flt files provide slightly better
positional precision, while for the long exposures, there is little difference
in precision.
Thus, we choose to use the $_{-}$flt versions of the ACS data sets.

Once again, we use the 2023 version of the
{\it hst1pass} code \citep{and2022}
to obtain positions and magnitudes of detected sources.
Note that there are standard PSF library files and distortion
corrections for both filters of ACS data sets being used here.
The positions thus obtained have undergone all
astrometric corrections implemented by {\it hst1pass},
including distortion and CTE correction. More specifically,
the distortion is based on the work done in
\citet{kp2015} and \citet{kp2018}. 
Preliminary tests indicated that CR-correction did not significantly affect
the performance of {\it hst1pass} detection or centering, for these data sets.
Thus, we chose not to precorrect for CRs in the ACS images.

As in Sec. \ref{subsec:wfpc2}, we perform plate
transformations for exposures taken at the same epoch and in the
same filter to assess the positional precision.
Long and short exposures are grouped separately.
Single-measurement standard errors are calculated for well-measured stars.
For filter F814W we obtain $\sim 15$ mpix in the long exposures
and $\sim 25$ mpix in the short exposures. The 2014- and 2021-epoch
data have comparable errors. For filter F475W we obtain
24 mpix. These errors correspond to 0.75 mas, 1.25 mas, and
1.2 mas respectively.
Comparing these values with the ones
obtained for WFPC2 in Sec \ref{subsec:wfpc2}, 
it is clear that ACS/WFC positions are more precise.
However, the time baseline of 22 years compared to
7 years with solely the ACS data renders the WFPC2
exposures quite competitive in the proper-motion
determination.
Also, the possibility of residual systematics inflating the
final proper-motion errors, above that estimated from the random centering
errors alone, argues for utilizing the longest time baseline available.

\section{Proper Motions \label{sec:pms}}
\subsection{Relative Proper Motions \label{subsec:relative}}

Each chip of each camera is treated as a separate unit, i.e., each
chip has its own, individual $(x,y)$ system. 
Astrometrically corrected pixel coordinates, obtained
in Secs. \ref{subsec:wfpc2} and \ref{subsec:acs},
are converted to equatorial coordinates, $(\alpha,\delta)$,
using the linear WCS information
in the header of the image fits files. 
Note that we do not depend on the WCS coefficients being very precise;
they are not.
Subsequent, higher-order transformations between chips will be made during the
proper-motion reduction, overriding these approximate, low-order terms.
However, working in a roughly common, equatorial system greatly facilitates
star matching between the numerous images.
Equatorial coordinates for all detections on each chip
are then gnomonically projected
into $(\xi, \eta)$ standard coordinates assuming a uniform tangent point,
which is taken to be the position of And\,III's center from \citet{mcconn2012},
specifically $(\alpha,\delta)$ = (8.89083, 36.49778) degrees.

Given the geometry of the various exposures' overlap (Fig. \ref{fig:overlap}),
we choose to construct two separate proper-motion catalogs,
corresponding to the two ACS chips.
As reference exposure, we adopt the
initial 2014 F814W data set.
The $(\xi, \eta)$ positions from all other chips/exposures are transformed 
into either of the chips of this exposure, using polynomial transformations
with up to 4th-order coefficients in each coordinate. 
The PC and WF2 chips of the WFPC2 are paired with ACS chip1,
and WF3 and WF4 are paired with ACS chip2.
An iterative
least-squares procedure is employed to refine both the polynomial coefficients
of each chip/exposure into the reference exposure, and every object's proper motion;
the initial iteration assumes zero proper motions for all reference stars.
The reference stars that are used to perform the transformations
and compute the polynomial coefficients
are predominantly And\,III stars.
The proper-motion system is thus relative, and since
the reference stars are dominated by And\,III members,
the system is roughly that of And\,III.

We calculate proper motions\footnote{Throughout the paper, $\mu_{\alpha}$
is actually $\mu_{\alpha}~ cos~\delta$, and as units for
the proper motions we will use both mas yr$^{-1}$ and $\mu$as yr$^{-1}$.}
for all objects that have a minimum
of 10 separate position measurements and a minimum of 7 years time baseline.
Proper motions are the slope of simple linear fits to each coordinate as a function of
time, removing the highest outlier if it deviates more than $2.5\sigma$ 
from the best-fit line, until no such outliers remain. 
Formal relative proper-motion uncertainties are calculated
from the scatter about this best-fit line.

\begin{figure}
    \centering
    \includegraphics[scale=0.38,angle=-90]{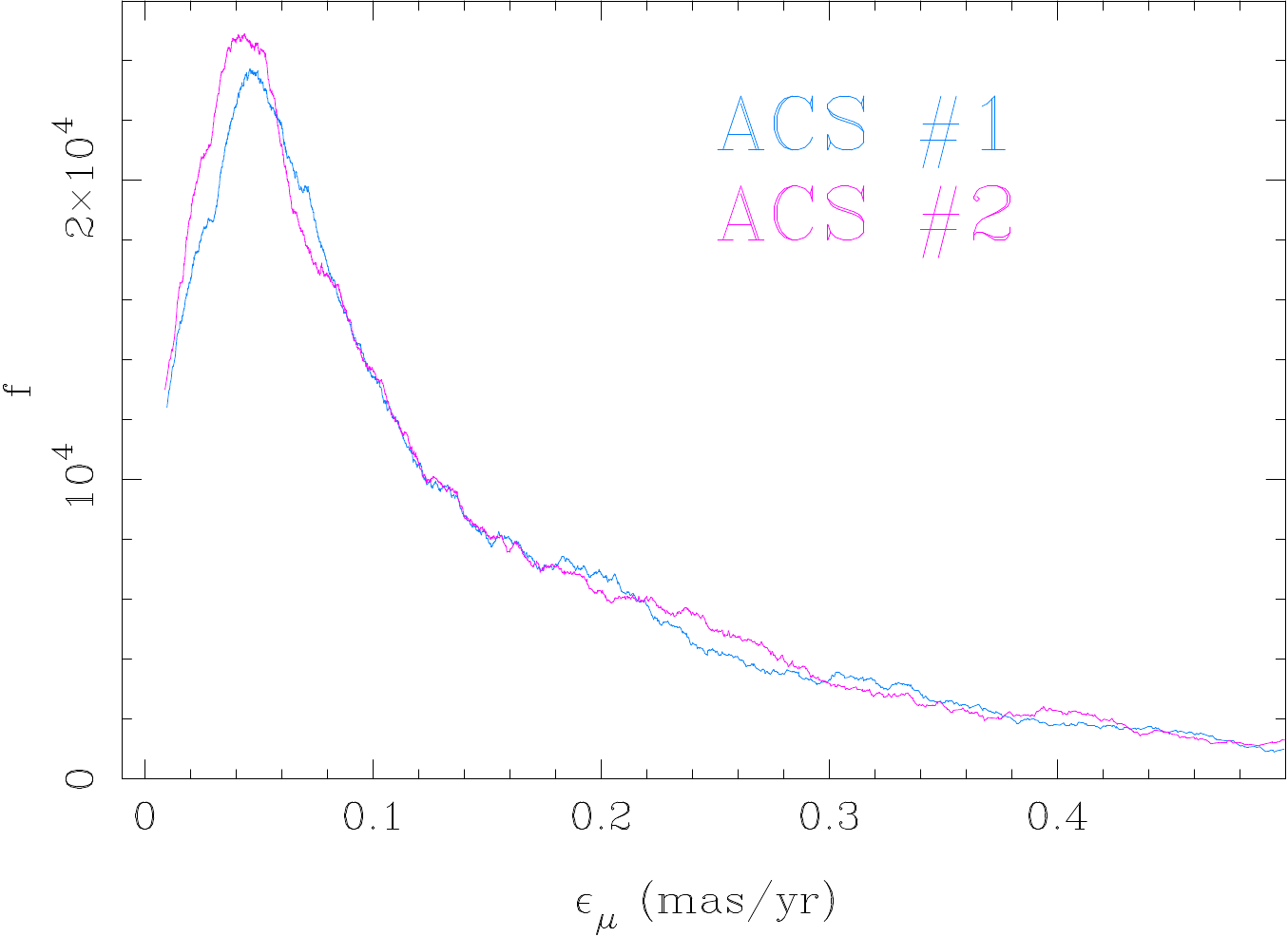}
    \caption{Distribution of formal proper-motion uncertainties in the two catalogs,
      corresponding to each ACS chip.
      Here we show uncertainties in $\mu_{\alpha}$; we obtain a similar distribution for
      $\mu_{\delta}$.}
    \label{fig:pmerror}
\end{figure}

In this manner, two {\bf independent} relative proper-motion catalogs
are constructed, one for each of the two ACS chips. 
The distribution of formal, random proper-motion uncertainties
in each catalog is shown in Figure \ref{fig:pmerror}. The
uncertainty distributions peak at $\sim 40~\mu$as yr$^{-1}$.

\begin{figure}
    \centering
    \includegraphics[scale=0.38,angle=-90]{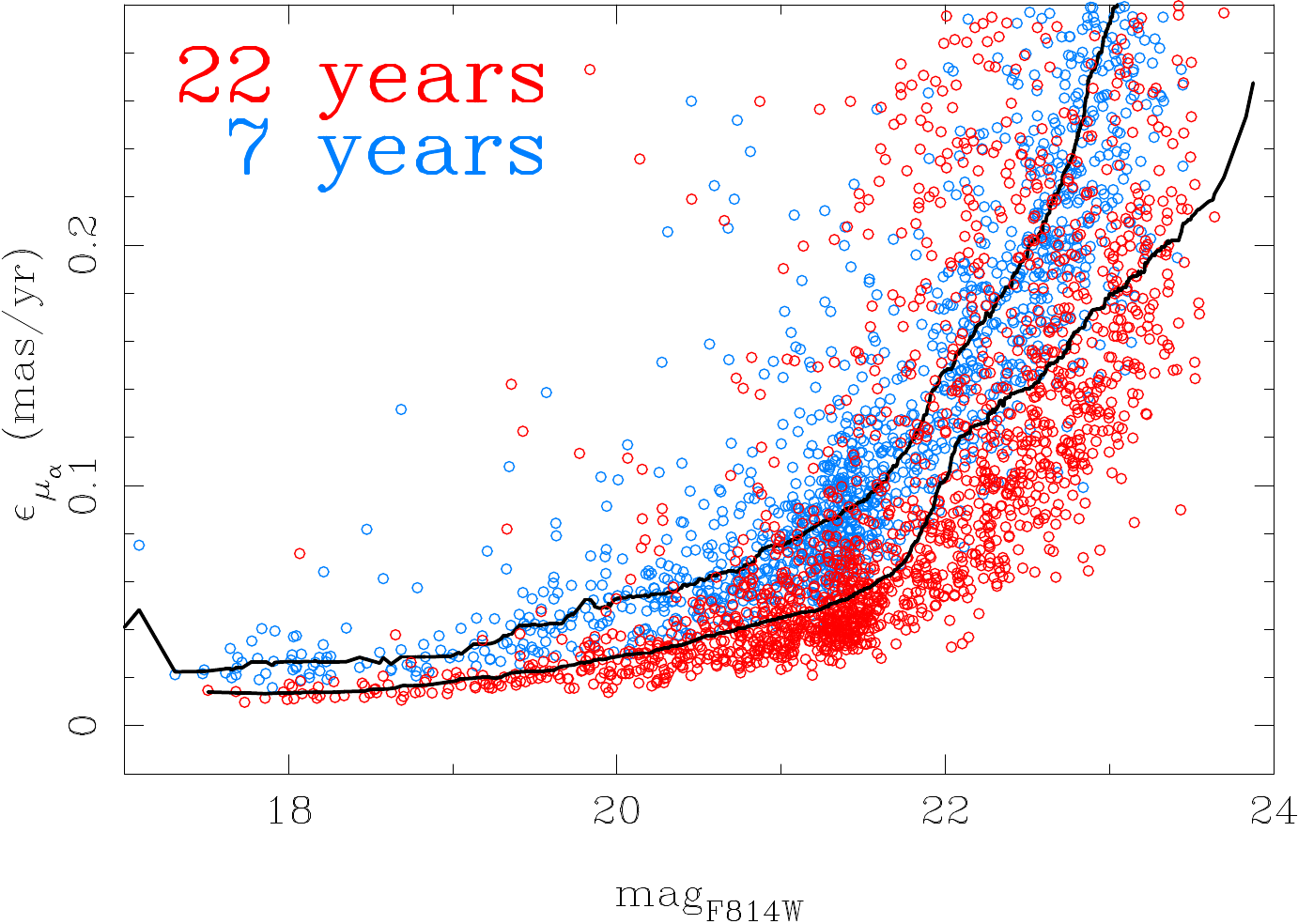}
    \caption{Formal proper-motion uncertainty as a function of
      instrumental magnitude in F814W for the ACS chip-1 catalog.
      Two groups are highlighted: stars that have a 7- and a 22-year baseline.
      A moving median is represented with a black line for each sample.
      For reference, the horizontal branch of And\,III is 
      at mag$_{F814W} \sim 21.4$, where a higher stellar density is apparent.}
    \label{fig:pmerr-mag}
\end{figure}

As one would expect, proper-motion uncertainties increase with magnitude,
as illustrated in Figure \ref{fig:pmerr-mag}.
Here, as an example, we show 
the proper-motion uncertainty in $\mu_{\alpha}$ as a
function of F814W instrumental magnitude
for the ACS chip-1 catalog.
(Similar looking plots are obtained for $\mu_{\delta}$
and for ACS chip 2.)
Stars with baselines of 7 years (ACS only) and 22 years (WFPC2 and ACS) are
highlighted. There are very few stars with a baseline of 15 years,
(i.e., missing ACS measures in 2021).
Formal proper-motion uncertainties of the best-measured stars are
of the order of $\sim 18~\mu$as yr$^{-1}$ for the 22-year sample.
These remain tightly distributed around a median value
out to mag$_{F814W} \sim 22$. Fainter than this value,
the uncertainties continue to increase in value and in scatter.
The 7-year sample, based on ACS data only, has best-measured uncertainties of
$\sim 25~\mu$as yr$^{-1}$ and exhibit a larger scatter compared to the
22-year sample. Toward faint magnitudes the 7-year uncertainties
increase, but with less scatter than that
of the 22-year sample.

\subsection{Correction to Absolute Proper Motions \label{subsec:absolute}}

We use {\it Gaia} Early Data Release 3 (EDR3) \citep{edr3}
stars and background galaxies to establish
the correction to absolute proper motion for our catalogs.
Values for these two types of correction are derived separately and later combined, by
weighted average, to achieve our final correction to absolute.
It will be seen that the EDR3 correction is significantly less certain than that based 
on galaxies, and contributes to the final value at a roughly 10 percent level.
Nonetheless, this correction anchors the bright end of our stellar proper motions
and is deemed an important check on the galaxies which anchor the faint end.
Furthermore, future {\it Gaia} Data Releases will improve the proper-motion
precision at the faint end. For example, it is expected that {\it Gaia} DR4 based on 66
months of observations (compared to 34 months in EDR3) will improve
precision by more than a factor of two. Thus the {\it Gaia}-based absolute correction can be
re-addressed once future releases are made available.

\subsubsection{EDR3-based correction}

EDR3 stars that appear in our catalogs are
at the faint end of {\it Gaia}'s reach and at the bright end of our study.
As such, EDR3 proper-motion errors will dominate the error
budget in the differences. It is worth noting that these EDR3 stars are
Galactic foreground
stars and therefore will have substantial proper motions.
Crucial to being able to include these stars are the
short ACS exposures taken in 2021
(see Tab. \ref{tab:prop-data}).

For the ACS chip-1 catalog we identify 8 EDR3 stars: all
with {\it Gaia} $G$ magnitudes between 17.1 and 20.8,
except for one star at $G=13.3$. This latter star
was effectively too bright for our study and thus we discard it
from consideration. The remaining 7 EDR3 stars
have EDR3 proper-motion uncertainties ranging from
0.06 to 1.16 mas yr$^{-1}$, while our
proper-motion uncertainty estimates range from 0.02 to 0.16 mas yr$^{-1}$.
For the ACS chip-2 catalog, we identify 10 EDR3 stars.
Again we eliminate the brightest one at $G=16.5$
which was poorly measured in our catalog as being
near saturation. The remaining 9 EDR3 stars
have $G$ magnitudes between 17.3 and 20.9.
Their EDR3 proper-motion uncertainties range between
0.06 and 1.46 mas yr$^{-1}$, while our catalog uncertainties
range between 0.01 and 0.18 mas yr$^{-1}$.
All 16 EDR3 stars have Renormalized Unit Weight Error
values between 0.93 and 1.04, and are thus
in the nominal range for single sources.
In our system of F814W instrumental magnitudes, the
EDR3 stars fall between mag$_{F814W} = 13$ and 17.

\begin{figure*}
    \centering
    \includegraphics[scale=0.70,angle=-90]{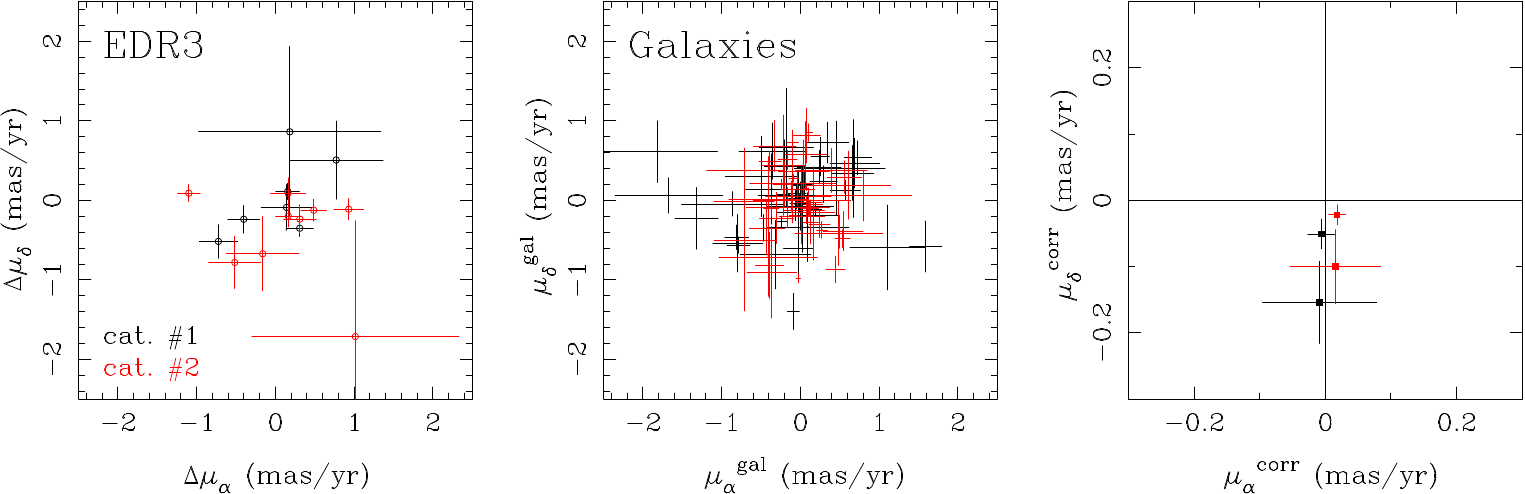}
    \caption{Correction to absolute proper motions as given by
      EDR3 stars (left panel) and galaxies (middle panel);
      each of the two ACS chip catalogs is as indicated.
      The right panel shows the weighted averages as computed
      in Tab. \ref{tab:cor}. The symbols with large error bars in this panel
      correspond to the EDR3-based values.
      Note the difference in scale
    between the first two panels and the right-most one.}
    \label{fig:abs-vpd}
\end{figure*}

We compute proper-motion differences between
our relative proper motions and EDR3 absolute ones.
The weighted mean of these differences is used
to determine the EDR3-based correction to
absolute proper motion for each catalog.
The weights are taken to be the quadrature sum of
the EDR3 proper-motion uncertainties and those from
our catalogs\footnote{An effort has been made to estimate the amplitude of possible
systematic errors for these bright stars in our sample.
Examining proper-motion trends with magnitude and extrapolating
to these stars' regime, tentative corrections in each field and 
along each axis vary but are only marginally significant.  Even
the largest of these has well under a 1-sigma effect on our
final combined zero-point correction, while the final combined
estimated uncertainties remain unchanged. For this reason, we
choose not to attempt to include them.
}.
The weighted-mean values are presented in Table
\ref{tab:cor} for the two catalogs, i.e., ACS chips. 
Even though the uncertainties of the
EDR3-based correction are somewhat large,
it is apparent that the correction is
near zero in $\mu_{\alpha}$ while it
significantly departs from zero in $\mu_{\delta}$.
In Figure \ref{fig:abs-vpd}, left panel we
show the distribution of these proper-motion differences,
where the error bars are the combined 1-sigma uncertainties, 
from EDR3 and our catalogs.

\subsubsection{Galaxies-based correction}

The 2014 F814W set of 22 exposures is used
to identify background galaxies in our two catalogs.
Each of these is CR-cleaned using the same
procedure as in Sec. \ref{subsec:wfpc2} for the WFPC2 images.
The ACS images are then shifted to align with the first one in the
set and median-combined to form a single, deep image.
This median image is searched by eye for galaxies.
In the chip-1 catalog we initially identify 65 galaxies, while
in the chip-2 catalog we have 83 galaxies with computed
relative proper motions. We then discard all
galaxies with combined proper-motion uncertainty
$\sqrt{(\epsilon_{\mu_{\alpha}}^2+\epsilon_{\mu_{\delta}}^2)} > 1.0$ mas yr$^{-1}$.
Galaxies with large proper motions (more than $3\sigma$ from the mean)
are also discarded.
We are thus left with 46 galaxies in the ACS chip-1 catalog,
and 52 galaxies in the ACS chip-2 catalog.

Our measures for the relative proper motions of these galaxies are shown in
Fig. \ref{fig:abs-vpd}, middle panel.
We calculate once again a weighted mean proper motion
of the galaxies, where the weights are given by the proper-motion
errors. These galaxy-based corrections to absolute proper motions
are also listed in Tab. \ref{tab:cor} and are to be
directly compared with the EDR3-based values in the same Table.
It is reassuring that the galaxy-based corrections also indicate
a value near zero in  $\mu_{\alpha}$ while showing a
significant departure from zero in $\mu_{\delta}$ and in the same
negative direction as the EDR3 correction.
The weighted averages for both galaxies and EDR3 correction are shown in the right panel of
Fig. \ref{fig:abs-vpd}.

\begin{deluxetable*}{lcrrcrr}[h]
  \tablecaption{Corrections to Absolute Proper Motion from EDR3 stars and galaxies
    \label{tab:cor}}
\tablewidth{0pt}
\tablehead{
  \colhead{Cat.} &
   \colhead{$N_{EDR3}$} &
    \colhead{$\mu_{\alpha}^{EDR3}$} &
    \colhead{$\mu_{\delta}^{EDR3}$} &
    \colhead{$N_{Gal}$} &
    \colhead{$\mu_{\alpha}^{Gal}$} &
    \colhead{$\mu_{\delta}^{Gal}$}  \\
    \colhead{} &
    \colhead{} &
    \colhead{(mas yr$^{-1}$)} &
    \colhead{(mas yr$^{-1}$)} &
    \colhead{} &
    \colhead{(mas yr$^{-1}$)} &
    \colhead{(mas yr$^{-1}$)} 
    }
\startdata
\#1 & 7 &  $-0.009\pm0.087$ & $-0.154\pm0.062$ & 46 & $-0.006\pm 0.021$&  $-0.051\pm 0.022$\\
\#2 & 9 &  $ 0.015\pm0.069$ & $-0.100\pm0.056$ & 52 & $0.018\pm 0.013$ & $-0.022\pm0.015$ \\
\enddata
\end{deluxetable*}

\section{And\,III's Absolute Proper Motion \label{sec:a3abs}}

Finally, to determine the systemic, absolute proper motion of And\,III
one must compute the average relative motion of And\,III stars and
subtract from it the correction to absolute proper motion.
Our adopted correction to absolute is a weighted mean of the two
values in Tab. \ref{tab:cor}, i.e., the EDR3- and galaxy-based corrections.
The final adopted corrections are listed in Table \ref{tab:a3all},
columns 2 and 3.

The vast majority of stars in our sample are members of And\,III and,
as such, can be used to calculate its systemic, relative motion.
We use stars with
mag$_{F814W}$ between 17 and 24,
with combined proper-motion uncertainties
$\le 0.5$ mas yr$^{-1}$, and with total
proper motion  $\le 0.5$ mas yr$^{-1}$.
The mean and its uncertainty are computed using probability plots
\citep{hamaker1978} 
with trimming of the extreme $10\%$
of both wings to eliminate the influence of potential outliers.
The method essentially fits the inner $80\%$ of the distribution to 
a shifted and scaled Gaussian to determine the mean and uncertainty.
We choose to use this mean and not one weighted by the proper-motion
uncertainties, as the latter leads to an unrealistically small
uncertainty in the weighted mean, of the order of $1~\mu$as yr$^{-1}$ .

In the chip-1 catalog there are 2396 stars
participating in the mean, and
in the chip-2 catalog there are 2176 stars.
The mean relative proper motion in each catalog
is listed in Table \ref{tab:a3all},
columns 4 and 5.

Note that we have also performed a mean relative motion based on And\,III
members selected via the satellite's sequence in the
color-magnitude diagram (CMD).
For this purpose a CMD was constructed from the 2014 data set.
Instrumental magnitudes in F814W and F475W are obtained as
averages of individual object measurements in each exposure.
Each exposure is placed on the ``system'' of the first exposure
in the 2014 data set by determining offsets between the
magnitudes in the reference exposure and the target one.
The CMD-based mean proper motion agrees within errors to the
magnitude-only trimmed star sample.
This confirms that, in the magnitude range we
considered, the stars in our field are
predominantly And\,III members.
Furthermore, tests demonstrate that the derived mean motion    
is {\it not} sensitive to the faint-magnitude limit adopted.

\begin{figure}[h]
    \centering
    \includegraphics[scale=0.50,angle=0]{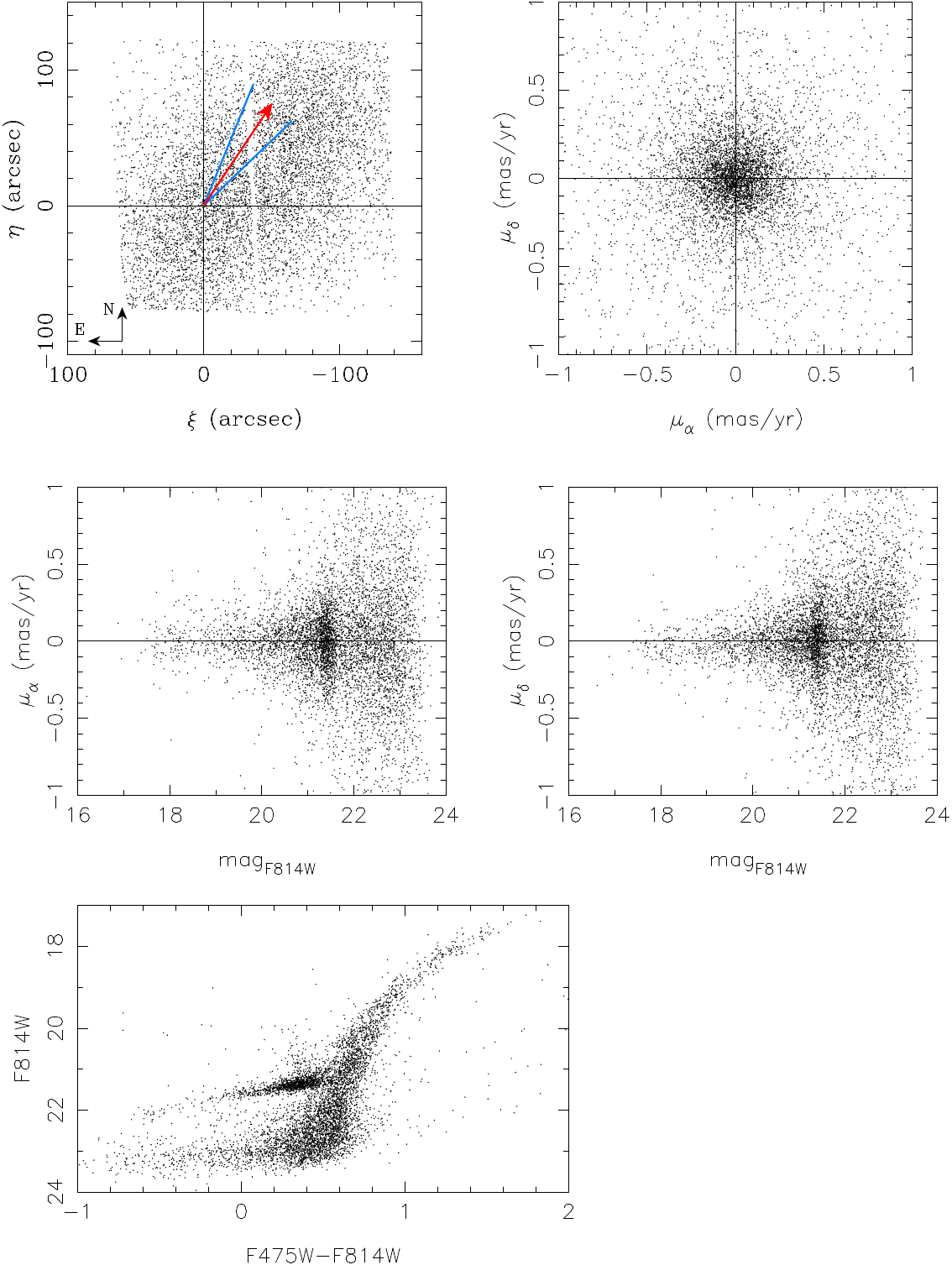}
    \caption{Spatial distribution, proper motions and CMD of
      our two catalogs, for stars with proper-motion uncertainties
      $\le 0.5$ mas yr$^{-1}$. The vector in the first panel marks the proper motion of And\,III
      with respect to M31 and its cone of uncertainty. M31's proper motion is
      the weighted average with the EDR3 proper motion from \citet{ps2021},
      see also Sec. \ref{subsec:orb-prop}.}
    \label{fig:vpd}
\end{figure}

In Figure \ref{fig:vpd} we show the characteristics of our two
proper-motion catalogs, for stars with proper-motion
errors $\le 0.5$ mas yr$^{-1}$. The upper-left panel shows the
spatial distribution of stars, where the oval shape of
the galaxy is apparent in the stellar density. Here, we also display
the vector of the proper motion of And\,III with respect to M31, and its cone of uncertainty.
For M31, we used the weighted average with the EDR3 proper motion from \citet{ps2021}
as detailed in Sec. \ref{subsec:orb-prop}. It appears that this motion
is aligned with the elongation of And\,III.
The upper-right
panel in Fig. \ref{fig:vpd}  shows the relative proper motion distribution,
followed by the run of proper motions with magnitude in the middle panels.
The concentration of stars at magnitude $\sim 21.5$ is the horizontal branch
of And\,III.
Finally, we present the instrumental CMD obtained from the 2014 data,
for stars with proper-motion uncertainties $\le 0.5$ mas yr$^{-1}$,
showing that the dominant stellar population is that of And\,III.

Finally, we note that a line-of-sight velocity membership selection
was made using data from \citet{kirby2020}. In the ACS chip-1 catalog
there were 13 stars with measured line-of-sight velocities,
while in the chip-2 catalog there was only one star.
These stars are very near the bright limit of our catalogs:
F814W instrumental mag = 17.6 - 19.3. Therefore we fear that
residual CTE is likely to affect the mean proper motion
at a level larger than the formal errors, for so few stars.
Thus, we have decided not to make use of this membership information.

The absolute proper-motion values obtained independently
per each ACS chip are listed in the last two columns of
Tab. \ref{tab:a3all}. Finally, the adopted
absolute proper motion of And\,III is taken to be the straight average of the
two determinations, with the uncertainty estimated from the difference of these values.
We believe the scatter (albeit only two determinations)
is a better real-world assessment of the uncertainty of our final measurement
as residual systematic errors are probably present at this level.
Our final value for the absolute proper motion of And\,III is thus
$(\mu_{\alpha} , \mu_{\delta}) = (-10.5\pm12.5, 47.5\pm12.5)~\mu$as yr$^{-1}$.

\begin{deluxetable*}{lrrrrrr}
  \tablecaption{And\,III: Relative and Absolute Proper Motion
    \label{tab:a3all}}
\tablewidth{0pt}
\tablehead{
  \colhead{Cat.} &
    \colhead{$\mu_{\alpha}^{cor}$} &
    \colhead{$\mu_{\delta}^{cor}$} &
    \colhead{$\mu_{\alpha}^{rel}$} &
    \colhead{$\mu_{\delta}^{rel}$} &
    \colhead{$\mu_{\delta}^{abs}$} &
    \colhead{$\mu_{\delta}^{abs}$}  \\
    \colhead{} &
    \colhead{(mas yr$^{-1}$)} &
    \colhead{(mas yr$^{-1}$)} &
    \colhead{(mas yr$^{-1}$)} &
    \colhead{(mas yr$^{-1}$)} &
    \colhead{(mas yr$^{-1}$)} &
    \colhead{(mas yr$^{-1}$)} 
    }
\startdata
\#1 &  $-0.006\pm0.020$ & $-0.063\pm0.021$ & $-0.004\pm 0.004$&  $-0.003\pm 0.003$ & $0.002\pm 0.020$&  $0.060\pm 0.021$ \\
\#2 &  $ 0.018\pm0.013$ & $-0.027\pm0.014$ & $-0.005\pm 0.005$ & $0.008\pm0.004$ & $-0.023\pm 0.014$ & $ 0.035\pm0.015$ \\
\enddata
\end{deluxetable*}

\begin{deluxetable*}{ccccccccc}
  \tablecaption{Orbital parameters.
    \label{tab:orbits}}
\tablewidth{0pt}
\tablehead{
  \colhead{M31 PM} &
  \colhead{$M_{\mathrm{vir,M31}}$} &
  \colhead{$f_{\mathrm{peri}}$\tablenotemark{a}} &
  \colhead{$t_{\mathrm{peri}}$\tablenotemark{b}} &
  \colhead{$r_{\mathrm{peri}}$\tablenotemark{c}} &
  \colhead{$f_{\mathrm{apo}}$\tablenotemark{a}} &
  \colhead{$t_{\mathrm{apo}}$\tablenotemark{b}} &
  \colhead{$r_{\mathrm{apo}}$\tablenotemark{c}} &
  \colhead{$e$\tablenotemark{d}} \\
  \colhead{} &
  \colhead{($\times10^{12}\,M_{\odot}$)} & 
  \colhead{(\%)} &
  \colhead{(Gyr)} &
  \colhead{(kpc)} &
  \colhead{(\%)} &
  \colhead{(Gyr)} &
  \colhead{(kpc)}
}
\startdata
HST+sats & 1.5 & 29 & -- $[2.4, 5.2]$ & -- $[49, 84]$ & 52 & 4.6 $[1.5, 4.6]$ & 490 $[184, 483]$ & 0.70 $[0.57, 0.86]$\\
" & 2.0 & 54 & 4.9 $[2.2, 5.0]$ & 76 $[49, 89]$ & 78 & 2.4 $[1.1, 3.9]$ & 305 $[164, 474]$ & 0.60 $[0.50, 0.80]$\\
\hline
HST+sats+EDR3 & 1.5 & 11 & -- $[3.3, 5.6]$ & -- $[53, 86]$ & 28 & -- $[2.0, 4.8]$ & -- $[245, 508]$ & 0.83 $[0.68, 0.89]$\\
" & 2.0 & 31 & -- $[2.8, 5.3]$ & -- $[49, 86]$ & 57 & 4.4 $[1.7, 4.5]$ & 525 $[217, 533]$ & 0.72 $[0.60, 0.85]$\\
\enddata
\tablecomments{Orbital parameters of And III after integrating backward for 6 Gyr. Uncertainties are included as $[15.9, 84.1]$ percentiles. Parameters denoted with a dash represent cases lacking a last pericentric or apocentric passage within the last 6 Gyr.}
\tablenotetext{a}{Fraction of orbits that achieved peri/apocentric passage over the last 6 Gyr.}
\tablenotetext{b}{Lookback time to last peri/apocentric passage.}
\tablenotetext{c}{Distance from M31 at the last peri/apocentric passage.}
\tablenotetext{d}{Eccentricity of the orbit.}
\end{deluxetable*}

\section{Orbit Analysis \label{sec:disc}}

\subsection{Orbital Properties and Constraints on M31's Mass \label{subsec:orb-prop}}

We calculate the space position and velocity of And\,III by combining our derived proper motion with other observed parameters.
Namely, we use RR Lyrae-based heliocentric distances for M31 ($776.2^{+22}_{-21}\,\mathrm{kpc}$) and And\,III ($721.1^{+17}_{-16}\,\mathrm{kpc}$) by \citet{savino2022}, along with radial velocities of $-300.1\pm3.9\,\mathrm{km}\,\mathrm{s}^{-1}$ and $-344.3\pm1.7\,\mathrm{km}\,\mathrm{s}^{-1}$ \citep{mcconn2012}, respectively.
In particular, the value we adopt for M31's proper motion strongly influences the resulting kinematics of And\,III.
We therefore test two PM estimates: the fiducial HST+Sats value used in \cite{sohn2020}
($\mu_\alpha ,\:\mu_\delta) = (34.3\pm8.4,\:-20.2\pm7.8)$~$\mu$as yr$^{-1}$ ,
as well as its weighted average with the EDR3 proper motion by \citet{salomon2021}.
The latter ($\mu_\alpha,\:\mu_\delta)=(40.1\pm6.6,\:-28.3\pm5.6)$~$\mu$as yr$^{-1}$
calculated in \citet{ps2021} arguably represents our current best estimate of M31's systemic proper motion.

To describe the space positions and velocities for And\,III, we adopt the M31-centric coordinate system as follows.
Observed parameters are first transformed into a Cartesian Galactocentric frame, then translated into M31's rest frame.
Finally, we apply a rotation such that the X-Y plane is aligned with M31's galactic disc, while the X-axis points away from the Milky Way's direction.
This is functionally equivalent to the frame adopted in \cite{sohn2020}, although we instead adopt updated distances from \citet{savino2022}.
Observational errors are accounted for by propagating uncertainties for distances, radial velocities and proper motions for both And\,III and M31 in a Monte Carlo fashion.
We thus generate 100 initial conditions for backward integration alongside the most likely observed parameters.

\begin{figure*}
    \centering
    \includegraphics[scale=0.50,angle=0]{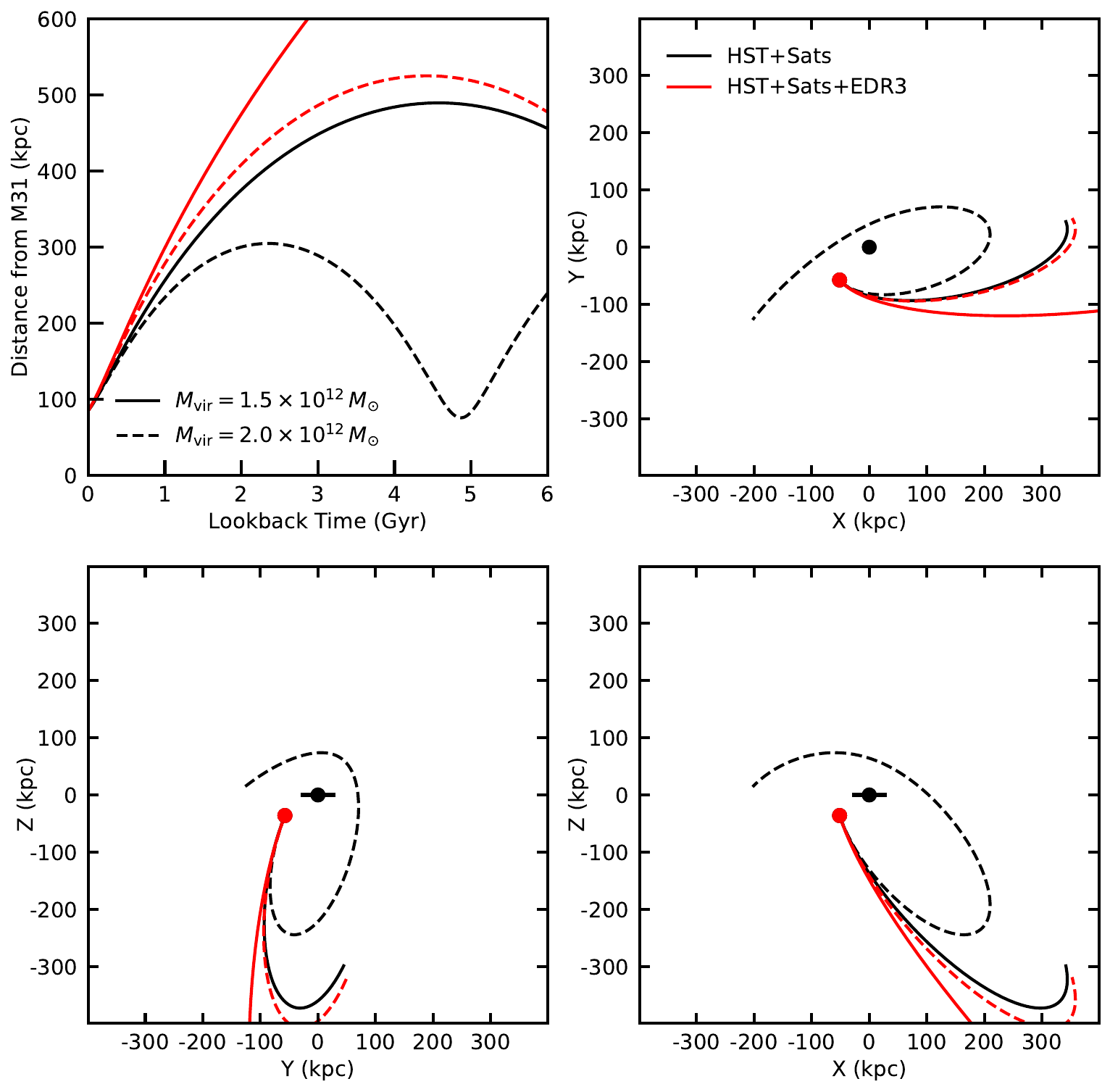}
    \caption{
    Backward orbit integrations of And\,III around M31 over the last 6 Gyr for a low-mass ($M_{\mathrm{vir}}=1.5\times10^{12}\,M_{\odot}$) and high-mass ($M_{\mathrm{vir}}=2\times10^{12}\,M_{\odot}$) M31 model, in solid and dashed lines, respectively.
    The black orbits adopt the HST+Sats M31 proper motion, while the red orbits assume a weighted average with M31's EDR3 proper motion.
    Andromeda's disc lies along the X-Y plane, while the Milky Way lies along the negative X-axis.
    }
    \label{fig:orbits}
\end{figure*}

To perform orbit integrations, we adopt two mass models for M31's potential from Table 2 of \citet{patel17} -- a higher-mass M31 with a virial mass of $M_{\mathrm{vir}}=2\times10^{12}\,M_{\odot}$ and a lower-mass M31 with $M_{\mathrm{vir}}=1.5\times10^{12}\,M_{\odot}$.
In addition, both potentials also contain a central Hernquist bulge and Miyamoto-Nagai disc to represent the baryonic galaxy.
Since we lack internal kinematics for And\,III, we estimate the effects of Chandrasekhar dynamical friction for a Plummer sphere with $M=10^{10}\,M_{\odot}$ and $R=1\,\mathrm{kpc}$, a model close to that adopted for NGC\,185 in \citet{sohn2020}.
The strength of the dynamical friction implemented thus serves as an upper bound (although we find modifying this component does not have a large impact on the resulting orbits).

We integrate And\,III and its Monte Carlo realizations backward for a period of 6 Gyr with M31's position fixed throughout.
Due to the individual error contributions in distance and proper motion for both And III and M31 especially, analytic orbits show a large scatter in their orbital properties (Table ~\ref{tab:orbits}).
In all realizations, And\,III is currently approaching pericenter and will perform its closest approach within the next $0.5\,\mathrm{Gyr}$ following a moderately eccentric orbit ($\left< e \right> = 0.6-0.83$).
Whether And\,III is bound or on first infall is uncertain and depends heavily on the adopted M31 mass and choice of proper motion.
Due to the direction of M31's motion enhancing the fly-by velocity of And\,III, the EDR3 PM weighted average generally requires a more massive M31 potential to keep And\,III bound compared to HST+Sats alone.
We note that And\,III in the higher-mass M31 potential with the EDR3-based proper motion for M31 follows a very similar orbit as in the lower-mass potential with HST+Sats, with its last apocenter around $4.5\,\mathrm{Gyr}$ ago.
If on first infall, And\,III would have been accreted from the far side of M31 and will spend a majority of its orbit there, marking its participation in M31's global asymmetric distribution of satellites towards the Milky Way \citep{savino2022} a purely transient event.

\begin{figure}[h]
    \centering
    \includegraphics[scale=0.50,angle=0]{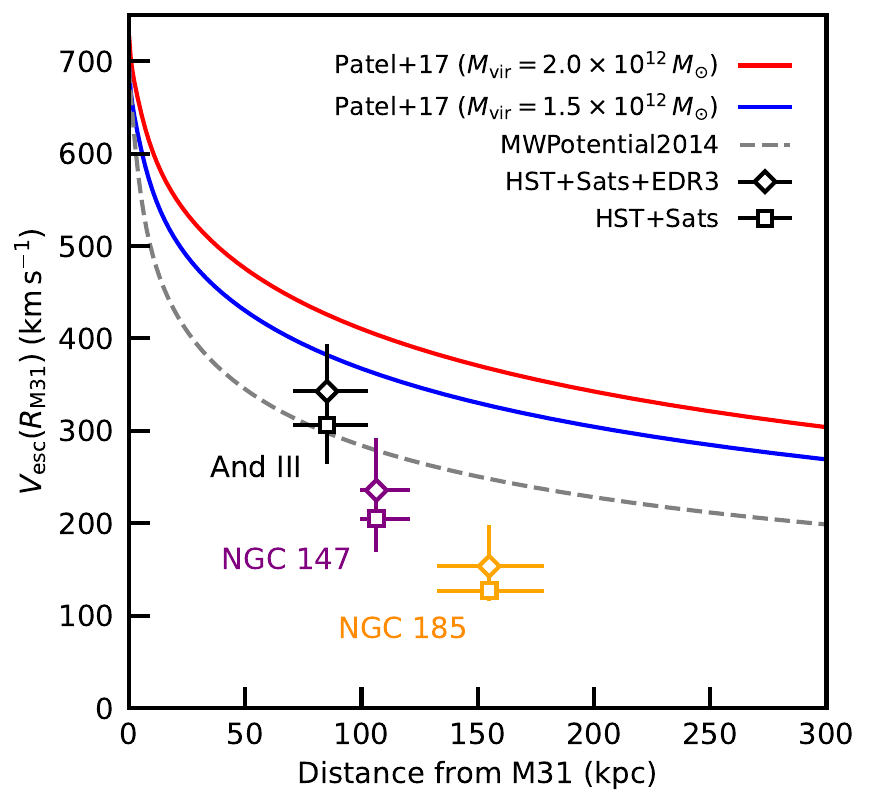}
    \caption{
    The M31-centric distance and total velocity of And\,III (black) using the HST+sats M31 proper motion (square) and its weighted average with EDR3 (diamond).
    Escape velocity curves of the two M31 mass models adopted in this paper are shown in blue and red, while a Milky Way-like potential is displayed for reference (grey dashed line).
    Results for NGC\,147 and NGC\,185, two other M31 satellites with proper motions, are also shown in purple and orange respectively.
   }
    \label{fig:escvel}
\end{figure}

Due to its higher M31-centric velocity compared to NGC\,147 and NGC\,185, the kinematics of And\,III serves as a useful probe to constrain the virial mass of M31 under the assumption that it is bound to M31, not on first infall.
In this case, the satellite's total velocity $v_{\mathrm{tot}}$ in the host's rest frame must not exceed the escape velocity $v_{\mathrm{esc}}=\sqrt{-2\,\phi(r)}$ (where $\phi(r)$ is the gravitational potential of the halo and any baryonic galaxy component), thus setting a lower bound on the permissible host mass.
In Fig.~\ref{fig:escvel}, we compare the total velocity of And\,III to the escape velocities at each radius from our two adopted mass models for M31 \citep{patel17}.
For both M31 proper motions, And\,III appears to be bound to M31 at $M_{\mathrm{vir}}\geq 1.5\times10^{12}\,M_{\odot}$.
And\,III is fully above the $v_{\mathrm{esc}}$ curve if M31 were to have a similar halo mass as the Milky Way, which effectively constrains M31's potential to be more massive than our Galaxy's, if And\,III is gravitationally bound.

\begin{figure}[h]
    \centering
    \includegraphics[scale=0.50,angle=0]{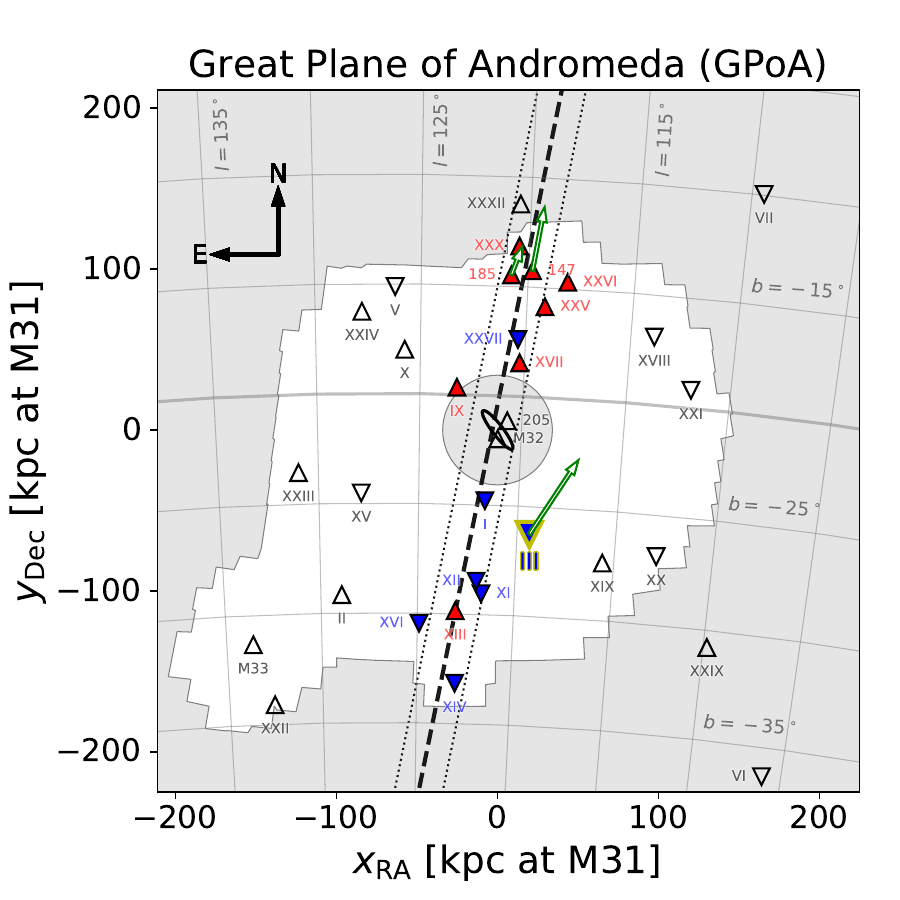}
    \caption{Spatial distribution of satellite galaxies
      around M31.
      Upward/downward triangles mark those satellites
   with receding/approaching line-of-sight velocities
   relative to that of M31. Filled
   symbols mark the candidate members of the GPoA according
   to their line-of-sight velocities and spatial
   distribution, with red/blue color also indicating the
   receding/approaching satellites.
   The GPoA and its estimated width
   are highlighted with dashed and dotted lines, respectively.
   The on-sky velocities relative to that of M31 for
   NGC\,147, NGC\,185 \citep{sohn2020} and And\,III (this work)
   are marked with green arrows.
   }
    \label{fig:vel-vect}
\end{figure}

\subsection{Alignment with to GPoA \label{subsec:alignment}}

\begin{figure*}[h]
    \centering
    \includegraphics[scale=0.50,angle=0]{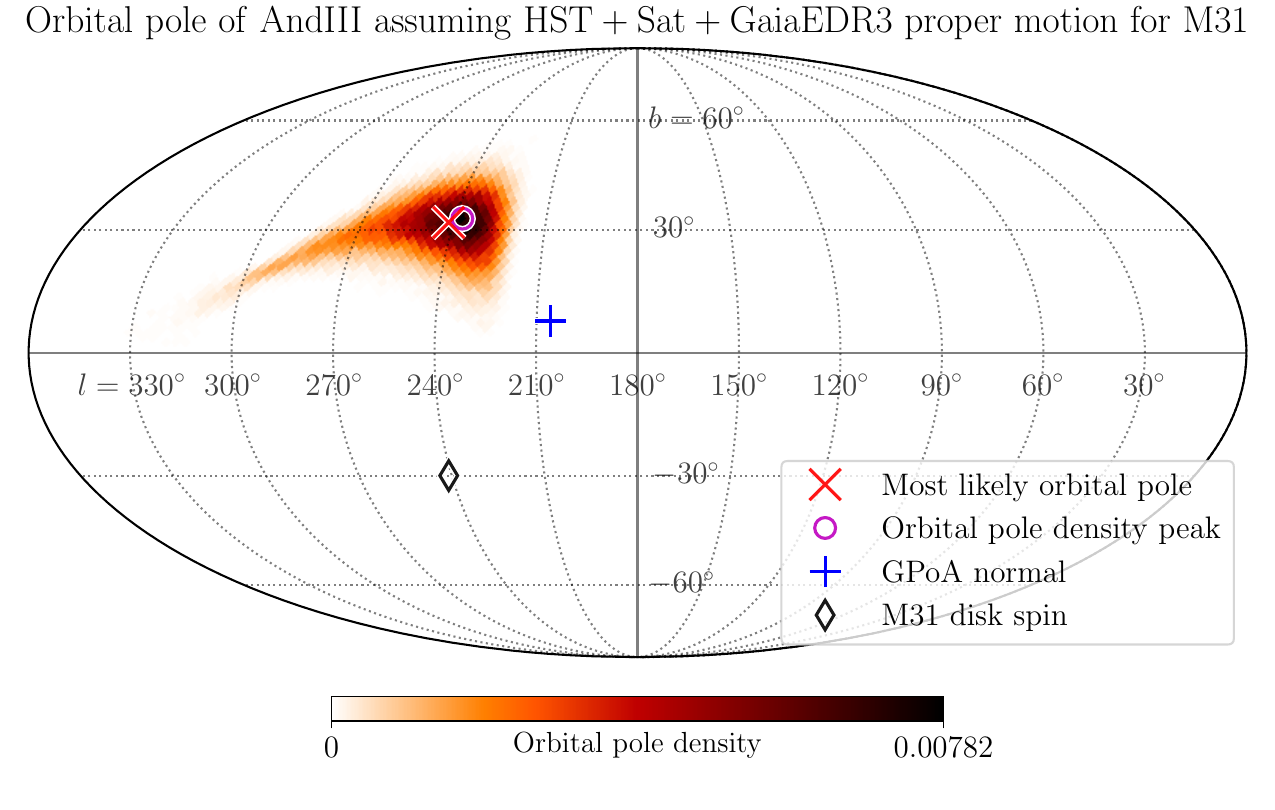}
    \caption{Distribution of calculated orbital poles of And\,III relative to M31 from $10^5$\ Monte-Carlo realizations, in Galactic coordinates. The orbital pole corresponding to the most likely measured position and velocity is shown as a red cross, the peak in the density distribution as a magenta circle. The blue plus sign indicates the normal vector to the GPoA that corresponding to the direction of co-rotation inferred from the line-of-sight velocity coherence, while the black diamond indicates the spin of the galactic disk of M31.
    See Figure 2 in \citet{ps2021} for equivalent plots for the two other M31 satellites with measured proper motions, NGC\,147 and NGC\,185.
   }
    \label{fig:orbpoles}
\end{figure*}

\begin{figure}[h]
    \centering
    \includegraphics[scale=0.50,angle=0]{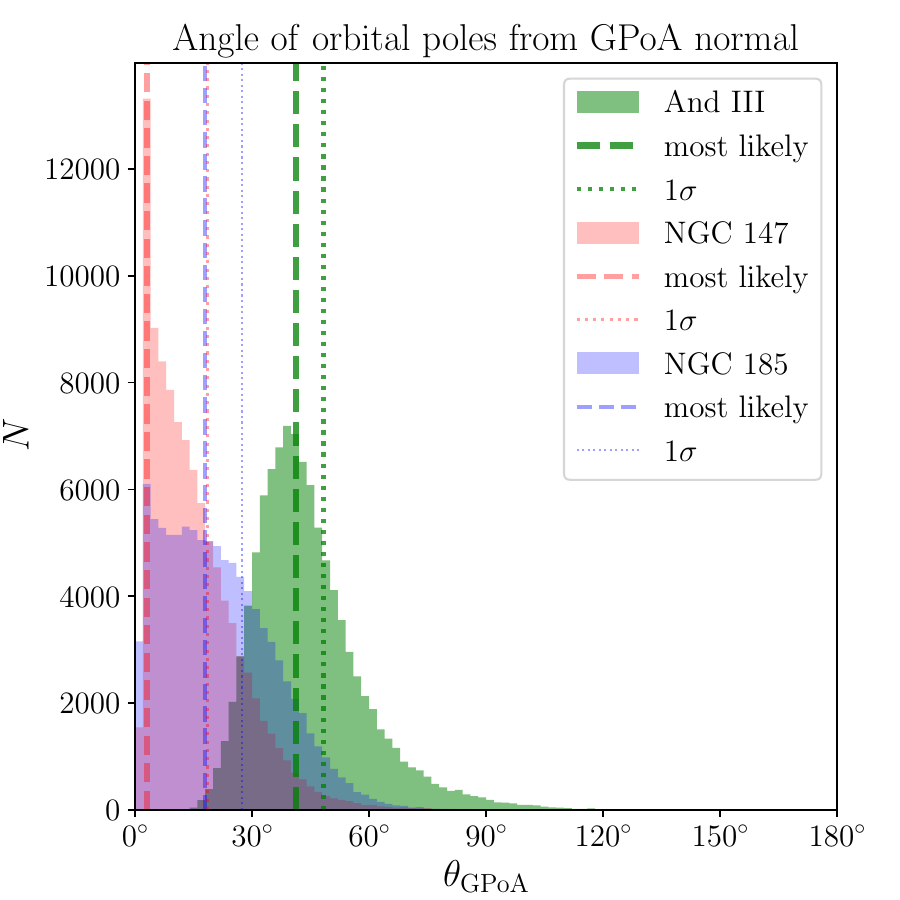}
    \caption{Distribution of the angle between the orbital pole of a satellite galaxy and the co-orbiting GPoA normal vector, based on $10^5$ Monte Carlo samplings from the measurement uncertainties. The dashed lines indicate the angle inferred from the most likely values, while the dotted lines indicate the $1\sigma$\ spreads. Since the spatial position of And\,III is offset by $21.5^\circ$\ from the GPoA, this is the minimum angular alignment that is geometrically achievable for its orbital pole.
   }
    \label{fig:GPoAangle}
\end{figure}

As discussed earlier, And\,III previously has been identified as a member of the Great Plane of Andromeda. It's spectroscopic systemic velocity follows the dominant line-of-sight velocity trend indicative of a co-rotating satellite plane. Two other on-plane satellite galaxies for which proper motions have been measured, NGC\,147 and NGC\,185, have both been found to be consistent with co-orbiting along this plane \citep{ps2021}.
In Figure \ref{fig:vel-vect} we show the on-sky distribution of M31's satellites with the velocity vectors
of the three satellites with measured proper motions highlighted.
Here, we repeat the analysis in \citet{ps2021} for And\,III.

Among satellites of M31, And\,III is the one most offset the GPoA (Fig. \ref {fig:vel-vect})
that is still considered a member of this structure. Its most-likely spatial position
already places it at an offset of $21.5^\circ$\ from the satellite plane.
This implies that its orbital pole can not align to better than this angle
with the normal to the satellite plane, for which we adopt $(l, b) = (205.8^\circ, 7.6^\circ)$\ following \citet{ps2021}.

As before, we consider two possible proper motions for M31. We then sample $10^5$\ Monte-Carlo realizations varying the positions, proper motions, and line-of-sight velocities of M31 and the satellite within their uncertainties. For each realization we measure the position of the orbital pole of And\,III relative to M31, and its angle to the GPoA normal vector. The resulting distribution of orbital pole directions is shown in Fig. \ref{fig:orbpoles}.
The distribution of alignment angles, in comparison to those inferred for NGC\,147 and NGC\,185, is shown in Fig. \ref{fig:GPoAangle}. Both figures are based on assuming the "HST+Sats+EDR3" proper motion for M31 (the plots look visually very similar if adopting the "HST+Sats" instead).

For the "HST+Sats" proper motion, we find that the peak of the orbital pole uncertainty cloud points to $(l, b) = (234.8^\circ, 30.0^\circ)$ and is $35.3^\circ$ from the GPoA normal, while the median offset of all realizations is $41.6^\circ$\ and has a standard deviation of $14.7^\circ$. The one (two) $\sigma$\ spread in this alignment angle is $30.4^\circ$ to $56.7^\circ$ ($21.3^\circ$ to $81.8^\circ$). Using only the most-likely values in all measured parameters, we find that the most-likely orbital pole points to $(l,b) = (241.2^\circ, 31.0^\circ)$\ and has an angular offset from the GPoA normal vector of $40.5^\circ$.

The alignment is slightly worse if we instead adopt the "HST+Sats+EDR3" proper motion. The orbital pole peak now points to $(l, b) = (237.7^\circ, 32.8^\circ)$, $38.8^\circ$ from the GPoA normal. The median offset is $42.1^\circ$\ with a standard deviation of $14.4^\circ$, while the one (two) $\sigma$\ spread is $31.9^\circ$ to $57.1^\circ$ ($23.5^\circ$ to $82.6^\circ$). The most-likely orbital pole points to $(l,b) = (241.8^\circ, 31.8^\circ)$, which implies an angular offset of $41.3^\circ$ from the GPoA normal.

Within the measurement uncertainties, And\,III appears to be consistent with co-orbiting along the GPoA to within the geometrically allowed limit. Its orbit is likely less well aligned than those of other two M31 satellites with measured proper motions, which are both consistent with a perfect orbital alignment, but Fig. \ref{fig:GPoAangle} shows considerable overlap in the distribution of possible orbital alignment angles of all three satellites, within a range comparable to the typically considered orbital pole alignment angle for the Milky Way satellite galaxies of $37^\circ$ \citep[see e.g.][]{taibi2024}. 
However, among the objects considered to  make up the M31 satellite plane, And\,III is the most spatially offset. As such, it could well have turned out to show an entirely unaligned orbit, or even one that is counter-orbiting (despite its agreement with the line-of-sight velocity trend, see its possible range of proper motions in Figure 11 of \citealt{ps2021}). This suggests that And\,III is indeed associated with the M31 satellite plane, though improved accuracy in the measured proper motion is required to firmly establish this.

\section{Summary \label{sec:sum}}

We have measured the absolute proper motion of And\,III using
ACS and WFPC2 exposures over three epochs and spanning up to 22 years.
Novel astrometric techniques have been used to process the WFPC2 images.
The spatial location of And\,III is some $20^{\circ}$ offset from the
GPoA satellite plane. Nevertheless, based on the line-of-sight velocity and
proximity to this plane, And\,III is considered a member of this plane.
The proper-motion measurement enables an orbit analysis that
favors dynamical membership to this structure. 
If And\,III is bound to M31, then it implies an M31 mass 
$M_{\mathrm{vir}}\geq 1.5\times10^{12}\,M_{\odot}$, larger than
that of the Milky Way. We also find that the motion of And\,III
with respect to M31 is aligned with the long axis of the satellite.

\newpage
\acknowledgments
This work was supported by program HST-AR-17029
provided by NASA through a grant from Space Telescope
Science Institute, which is operated by the
Association of Universities for Research in Astronomy, Inc.
MSP acknowledges funding via a Leibniz-Junior Research Group (project number J94/2020).

This study has made use of data from the European Space Agency
(ESA) mission {\it Gaia} 
(\url{https://www.cosmos.esa.int/gaia}),
processed by the Gaia Data Processing and Analysis Consortium (DPAC, 
\url{https://www.cosmos.esa.int/web/gaia/dpac/consortium}).
Funding for the DPAC has been provided by national institutions,
in particular the institutions participating in
the {\it Gaia} Multilateral Agreement.

All the {\it HST} data sets used in this paper can be found in MAST.
Set 1 corresponds to all WFPC2 observations, while Set 2 to all ACS observations. \\
Set 1: \dataset[http://dx.doi.org/10.17909/r78a-kn48]{http://dx.doi.org/10.17909/r78a-kn48} \\
Set 2: \dataset[http://dx.doi.org/10.17909/62eg-0f60]{http://dx.doi.org/10.17909/62eg-0f60} \\

\vspace{5mm}
\facilities{{\it HST}, MAST, {\it Gaia}}


\bibliography{ms}{}

\end{document}